\documentclass[iop,revtex4]{emulateapj}

\newcommand{\mytilde}{\raise.19ex\hbox{$\scriptstyle\sim$}}
\newcommand{\MOO}{MOO~J1014+0038}
\shorttitle{ICL STUDY OF MOO J1014+0038}
\shortauthors{Ko \& Jee}

\begin{document}

\title{Evidence for the existence of abundant intracluster light at $z= 1.24$}

\author{Jongwan Ko\altaffilmark{1,2} and M. James Jee\altaffilmark{3,4}}

\altaffiltext{1}{Korea Astronomy and Space Science Institute, Daedeokdae-ro, Daejeon 34055, Korea}
\altaffiltext{2}{University of Science and Technology, Gajeong-ro, Daejeon 34113, Korea}
\altaffiltext{3}{Department of Astronomy, Yonsei University, 50 Yonsei-ro, Seoul 03722, Korea}
\altaffiltext{4}{Department of Physics, University of California, Davis, One Shields Avenue, Davis, CA 95616, USA}

\email{jwko@kasi.re.kr; mkjee@yonsei.ac.kr}

\keywords{
galaxies: clusters: individual (MOO J1014+0038) ---
galaxies: high-redshift ---
galaxies: elliptical and lenticular, cD ---
galaxies: evolution
}

\begin{abstract}
Intracluster stars are believed to be unbound from their progenitor galaxies and diffused throughout the galaxy cluster, creating intracluster light (ICL). However, when and how these stars form are still in debate. To directly constrain the origin, one powerful method is to study clusters at the epoch when mature galaxy clusters began to appear. We report measurements of the spatial distribution, color, and quantity of diffuse intracluster stars for a massive galaxy cluster at a redshift of 1.24. This is the most distant galaxy cluster to date for which those three properties of the ICL have been quantified simultaneously. Our detection of the significant ICL fraction in this unprecedentedly high redshift regime strongly indicates that intracluster stars, contrary to most previous studies, might have formed during a short period and early in the history of the Virgo-like massive cluster formation and might be concurrent with the formation of the brightest cluster galaxy.
\end{abstract}

\section{Introduction}
Not all stars in the universe are gravitationally bound to galaxies. Since first discovered in 1951 (Zwicky 1951), observations have clearly revealed that a significant stellar component fills the space between galaxies in nearby galaxy clusters (e.g., Gregg \& West 1998; Feldmeier et al. 2002; Lin \& Mohr 2004; Gonzalez et al. 2005; Zibetti et al. 2005; Mihos et al. 2017), observed as intracluster light (ICL). Although several scenarios for the production of intracluster stars have been suggested by numerical simulations (e.g., Murante et al. 2007; Conroy et al. 2007; Puchwein et al. 2010), including in-situ formation, tidal disruption or/and stripping of galaxies, and the brightest cluster galaxy (BCG) formation, the dominant production mechanism is still in dispute. However, there is a consensus that most intracluster stars in local galaxy clusters are several billion years-long accumulated material since $z\mytilde$1 (Rudick et al. 2011; Contini et al. 2014; Cooper et al. 2015).

Observations of multiple tidal features in the Virgo cluster of galaxies suggest that perhaps a significant fraction of its intracluster stars might have been stripped from the cluster galaxies through numerous galaxy interactions (Mihos et al. 2017). Within this paradigm, observed properties of these stripped stars should reflect the history of the galaxy mergers and interactions. One obvious prediction if one accepts this ICL production mechanism as the dominant source is considerable evolution of ICL properties with redshift; for example, more evolved clusters should have a higher ICL fraction.

However, observational constraints on the origin of ICL have been limited because the typical surface brightness of ICL is extremely low, being less than $\mytilde$1\% of the night sky from the ground (Mihos et al. 2017). Thus, studies of detailed ICL features including identification of the distinct ICL populations, such as planetary nebulae and globular clusters, are possible so far for nearby clusters (Gerhard et al. 2007; Lee et al. 2010; Peng et al. 2011; Ko et al. 2017). Because the intracluster stars in nearby clusters have been stripped at different epochs, interpretation of the observation in the context of the ICL origin is difficult.

Beyond Virgo, current ICL observations have been limited to galaxy clusters mostly at $z <$ 0.6 (e.g., Jee 2010; DeMaio et al. 2015; Morishita et al. 2017) and marginally extended to $z\mytilde$0.9 (Guennou et al. 2012; Burke et al. 2015; DeMaio et al. 2018). Some observations show no strong variation of the ICL fraction with redshift (Krick \& Bernstein 2007; Presotto et al. 2014), which is contradictory to the scenarios suggested by recent simulations (Rudick et al. 2011; Contini et al. 2014; Cooper et al. 2015), wherein most intracluster stars are produced after $z\mytilde$1.

To directly constrain the origin, we must extend the redshift baseline significantly. Arguably, the most interesting epoch is between $z\mytilde$1 and 2, when the first mature galaxy clusters began to appear (e.g., Gobat et al. 2011; Muzzin et al. 2013; Prichard et al. 2017; Nantais et al. 2017); if ongoing stripping processes are indeed dominant,  young galaxy clusters in this redshift regime will have a much lower ICL fraction than similarly massive clusters in today's universe.

Here we report measurements of the radial surface brightness profile, color, and fraction of the ICL to the total cluster light for the massive galaxy cluster MOO J1014+0038 at $z =$ 1.24. This is the first cluster for which the surface brightness and color profiles of the ICL are measured at $z >$ 1. To date, the most distant clusters for which both the color and surface brightness profiles of the ICL have been measured simultaneously are at $z <$ 0.9 (DeMaio et al. 2018).

MOO~J1014+0038 was discovered in the Massive and Distant Clusters of Wide-field Infrared Survey Explorer (Wright et al. 2010) Survey (MaDCoWS; Gettings et al. 2012). The $M_{200}$ value inferred from the Sunyaev-Zel'dovich data (Brodwin et al. 2015) is $\mytilde5.6 \times 10^{14} M_{\sun}$ comparable to that of Virgo. The cluster was selected for the ``See Change" project (PI. S. Perlmutter), which is a large {\it Hubble Space Telescope} (HST) multi-epoch (Cycles 22 and 23) program to efficiently detect Type Ia supernovae from massive clusters at $z >$ 1 and robustly determine
high-redshift massive clusters via weak lensing (Rubin et al. 2017; Jee et al. 2017). We utilize the deep near infrared imaging data [Wide Field Camera 3 (WFC3) F105W and F140W] of MOO~J1014+0038 to characterize the ICL. In general, ICL studies with HST is hampered by the small field of view and thus the difficulty in reliably determining the background level.
However, for high-redshift clusters such as MOO~J1014+0038 at $z=1.24$, this weakness is greatly reduced not only because the plate scale at the redshift of the clusters is large, but also because the contamination by the ICL surface brightness is expected to be very low compared to that of low-redshift clusters.

We organize our paper in the following way. In~\textsection\ref{section_observations} we describe our data and basic reduction. Our ICL analysis methods and results are presented in \textsection\ref{section_analysis} and \textsection\ref{section_results}, respectively, before our conclusion in \textsection\ref{section_conclusions}. We adopt the cosmological parameters presented in Planck Collaboration et al. (2016), which give a plate scale of $\mytilde 8.55$ kpc/ $\arcsec$ at the redshift ($z=1.24$) of MOO~J1014+0038.

\section{Observations} \label{section_observations}
MOO~J1014+0038 has been observed as part of the ``See Change" programs (13677 and 14327) in Cycles 22 and 23. The integrated exposure times are approximately 17358~s, 16812~s, and 7448~s for the WFC3-IR F105W, F140W, and WFC3-UVIS F814W filters, respectively. Roll angles are rotated between different visits. The total exposure for this target is considerably higher than those for other clusters in ``See Change" because additional orbits were assigned to follow up the discovery of the lensed background supernova at $z=2.22$ (Rubin et al. 2017).

Our data reduction begins with the {\tt FLT} images output by the Space Telescope Science Institute (STScI) {\tt calwf3} pipeline. These images are flatfielded
with the default composite flats, which are derived by combining the ground-based high-frequency P-flats with residual sky flats. The claimed accuracy of the STScI pipeline flat is 0.5\% or less except for the region within 128 pixels of the detector edge (Gennaro et al. 2018). We independently verified this claim utilizing large HST survey data and the results are presented in \textsection\ref{section_flat_check}. Since this level of flat accuracy is sufficient for the scientific scope of the current paper, we do not attempt to further improve it.

The next step is to create robust stacks after carefully aligning exposures from different visits. Using common astronomical objects, we find that the typical relative astrometric error in the World Coordinate System (WCS) information of the MOO~J1014+0038 image headers is $\mytilde4\arcsec$. We create a shift file based on this measurement and run MultiDrizzle to create final stacks. 

A critical step in image stacking is consistent sky subtraction across all contributing exposures. The default behavior of MultiDrizzle is to determine a global sky level exposure by exposure 
by estimating the mean of the statistical distribution of pixels in the image with sigma-clipping and subtract this value from individual frames \footnote{This subtraction is performed when individually drizzled-images are created. The input file is not affected.}. Because of the small field of view of the detector, the sky level is likely to be overestimated if the ICL level is not negligible.
This  ``removal" of ICL, however, is not a concern as long as we later separately measure the background level again from the final stack, against which the ICL is quantified by determining the excess surface brightness above this baseline value. Of course, the key requirement, nevertheless, is to maintain consistency in the sky level determination for every exposure. For example, given two exposures of different dithers, their global sky values are not identical simply because the statistics are obtained from different regions of the sky. Therefore, as a scheme to use an
identical region for sky estimation, we define a common annulus centered on the BCG and use the pixels belonging to this annulus to estimate the sky level for each exposure. We save this sky value in the image header of the {\tt FLT} file and let MultiDrizzle use it for sky subtraction.

\section{Analysis} \label{section_analysis}

\subsection{Flat Verification} \label{section_flat_check}
Residual flat-fielding inaccuracy on a large scale is in general one of the important factors limiting interpretation of diffuse light measurements.
Since MOO~J1014+0038 is imaged with rotated roll angles, one can argue that the residual systematics in the azimuthal direction is greatly reduced compared to the aforementioned $\mytilde$0.5\% level. However, obviously the field rotation does not diminish any residual systematic errors in the radial direction.
\begin{figure*}
\centering
\includegraphics[width=8.5cm]{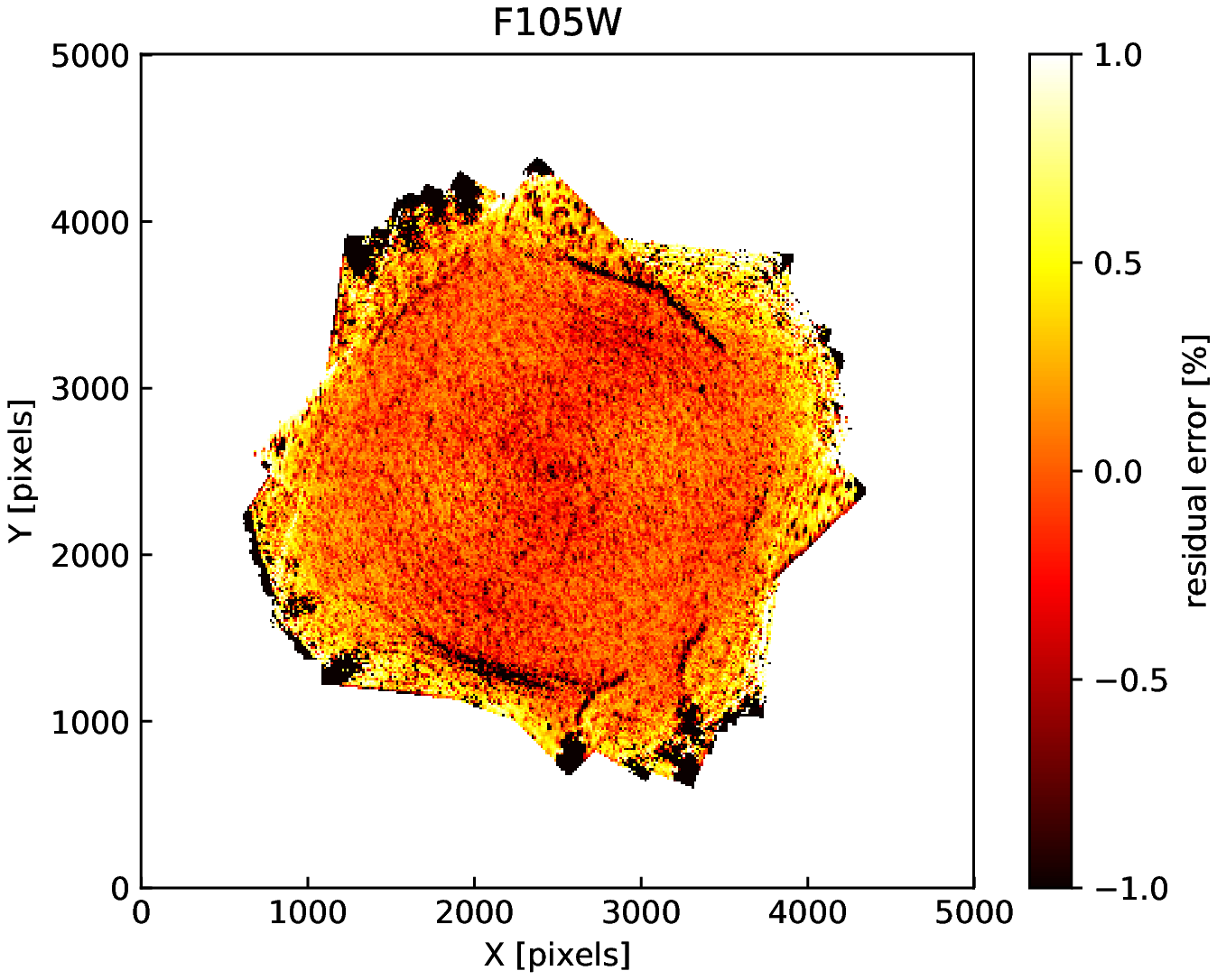}
\includegraphics[width=8.5cm]{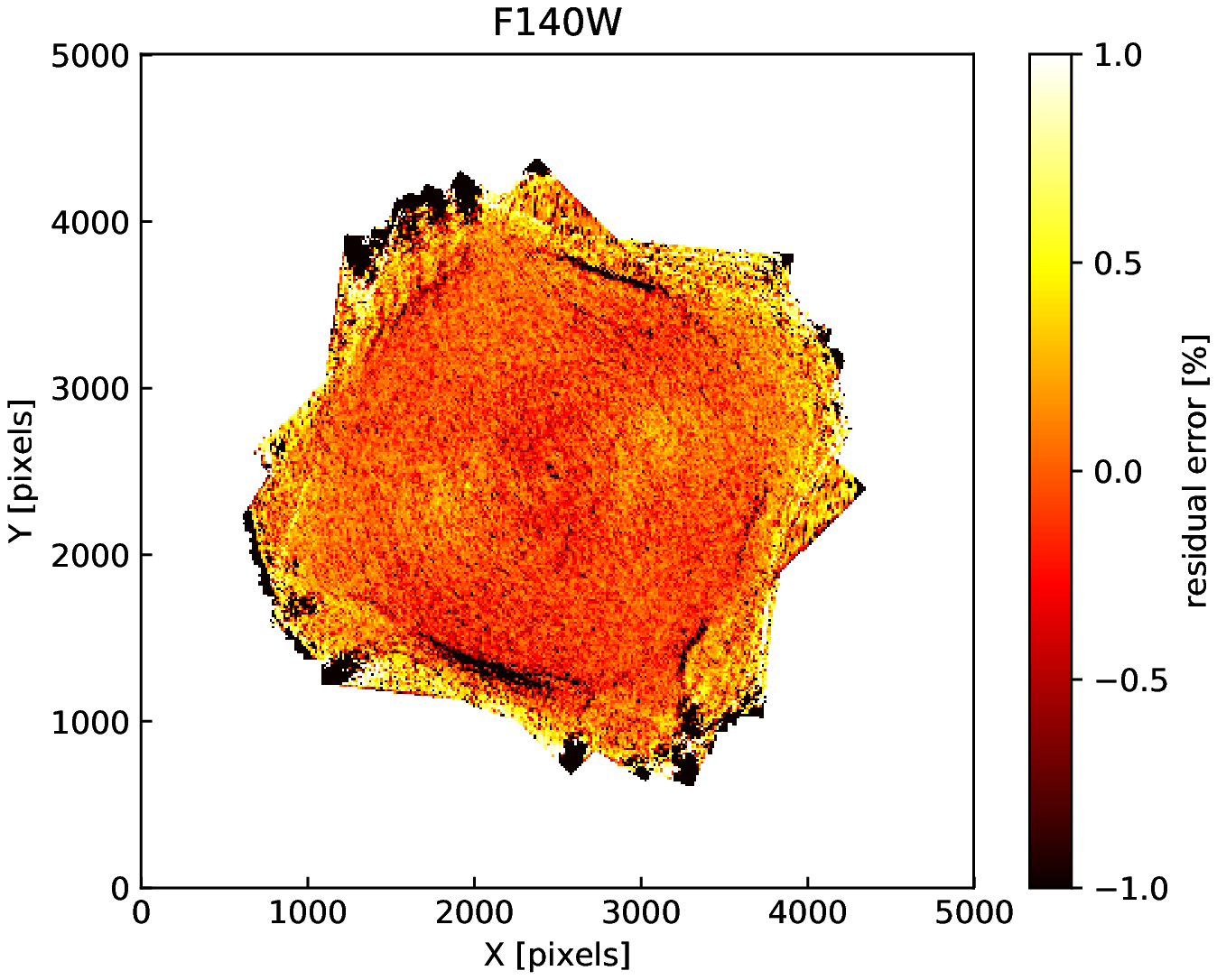}
\caption{Residual flat errors. We estimate the impact of the residual flat error on the coadd images by applying the same shifts and rotations applied to our science frames while ``drizzling". The residual flat errors in the central region [$r\lesssim1200$ pixels (1\arcmin)], where we measure ICL, are very small ($\mytilde0.2$\% or less). }
\label{fig:residual_flat}
\end{figure*}

\begin{figure}
\centering
\includegraphics[width=8.5cm]{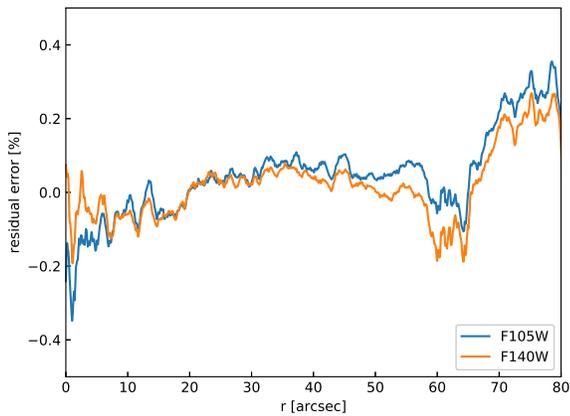}
\caption{Radial profile of the residual flat errors shown in Figure~\ref{fig:residual_flat}. We choose the center of the coadd image as the origin and compute azimuthally averaged residual flat errors. At $r\lesssim60\arcsec$, the residual errors are less than 0.2\% and will have a negligible impact on the current ICL measurement. }
\label{fig:residual_flat_profile}
\end{figure}

We independently verify the STScI flat accuracy by producing our own residual sky flats in the following steps. First, we retrieved WFC3 F105W and F140W FLT images from large survey programs such as Cosmic Assembly Near-infrared Deep Extragalactic Legacy Survey (CANDELS), WFC3 Infrared Spectroscopic Parallel (WISP), All-wavelength Extended Groth strip International Survey (AEGIS), etc. It is important to avoid frames from object-targeted observations because they often place objects in the detector center, which would cause bias in flat estimation. For each frame, we masked out objects and normalized the image using the median value. Finally, we median-stacked all FLT files. Because the FLT files are the results after the STScI pipeline application, our median-stacked images show residual flat-fielding errors, which indeed show $\mytilde$0.5\% level variations. As mentioned above, this systematics does not directly translate to our ICL measurement uncertainties because our final science image is created after combining multiple epoch data with different shifts and orientations using Multi-drizzle. We simulated this effect by replacing the cluster FLT images with our residual flat images and running Multidrizzle in the same way as we create the scientific image.
The outcome produced in this way generates the residual flat image reflecting different shifts and orientations of the input frames as shown in Figure~\ref{fig:residual_flat}. We found that the residual azimuthal variation is negligibly small ($<$0.1\%) thanks to the various roll angle rotations. The residual radial variation is present at the $\mytilde$0.2\% level (Figure~\ref{fig:residual_flat_profile}). Although small in contribution to the total error budget, we include this residual flat-fielding error in our ICL analysis.

\subsection{Background Level Determination}
In general, uncertainties in ICL measurements are dominated by background level estimation and flat-fielding inaccuracy. In the central region near the BCG, blending of the light from the outer regions of the galaxies with the ICL makes this analysis particularly challenging.

To determine our background level for the ICL measurements, the masked image (presented in \textsection\ref{section_masking}) is azimuthally divided into 24 wedges in three radial bins ranging from 42$\arcsec$ to 60$\arcsec$ (350$-$500 kpc), which gives a total of 72 sky bins. This scheme is similar to the method adopted in previous studies (e.g., DeMaio et al. 2015), but is different in that we add radial bins to take into account the radial variation of the background level. The inner radius is determined by locating the region where the surface brightness profile starts to flatten whereas the outer radius is determined by locating the region, where we can draw a complete circle without crossing image edges; our estimation shows that the residual
flat errors are also small ($\lesssim0.2$\%) within $r\lesssim60\arcsec$ (see \textsection\ref{section_flat_check}).
For each sky bin, we compute a 3$\sigma$-clipped median of the unmasked pixels with 10 iterations. The final background level is estimated by weight-averaging these 3$\sigma$-clipped median values from all 72 sky bins. We adopt the standard deviation as the background level errors, which we find to be 3.2 (29.1 mag arcsec$^{-2}$) and 4.2 ADUs (29.0 mag arcsec$^{-2}$) for F105W and F140W, respectively. Both errors correspond to $\mytilde$0.9\% of the mean sky level. These 1$\sigma$ background level ($\sigma_{sky}$) uncertainties are adopted as the detection limit of the ICL in the current analysis.

\begin{figure}
\centering
\includegraphics[width=8cm]{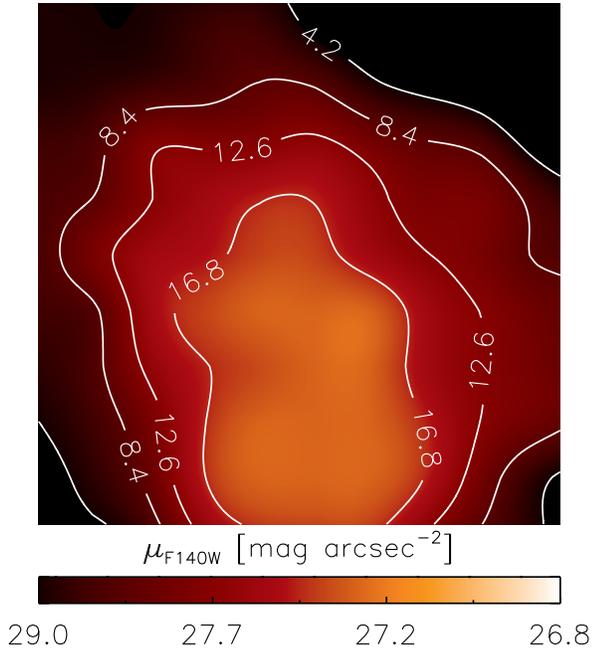}
\caption{WFC F140W background map for the central $1\arcmin\times1\arcmin$ region of the cluster, where we measure ICL. We produce this background map for object detection via SExtractor with a mesh size of 128 $\times$ 128 pixel. White contours show the 1, 2, 3, and 4$\sigma_{sky}$ (4.2$-$16.8 ADU) levels, where $\sigma_{sky}$ is the uncertainty of the background level. Color depicts the range of the background level in units of surface brightness, whose lower and upper limits correspond to 1$\sigma_{sky}$ and $7\sigma_{sky}$, respectively.}
\label{fig:back_map}
\end{figure}

If ICL is extended to a few hundred kpcs from the BCG, our background level determination would certainly be influenced by its presence. In Figure~\ref{fig:back_map}, we show the spatially varying local background map with contours showing the 1$-$4$\sigma_{sky}$ (4.2$-$16.8 ADUs corresponding to 29.0$-$27.4 mag arcsec$^{-2}$) levels, where $\sigma_{sky}$ is the uncertainty of the (global) background level. We produced the result via SExtractor (Bertin \& Arnouts 1996) with a mesh size of 128 pixels, which is small enough to measure the small scale variation of the local background level and large enough to be insensitive to the presence of objects. We verify that the SExtractor background level within the interval 42$\arcsec <r< 60\arcsec$ is highly consistent with the value determined from the above 72 sky bins. However, this method (i.e., estimation based on many sky bins) provides a conservative way to estimate the background level uncertainty.

\subsection{Object Detection and Masking} \label{section_masking}
Separating ICL components from the contribution of luminous astronomical objects (e.g., stars and galaxies) is a challenging task since the extended light from these sources smoothly blends into the ICL. In the current study, objects are detected with SExtractor by looking for three or more connected pixels ({\tt DETECT\_MINAREA=3}) one sigma above the local sky level ({\tt DETECT\_THRESH=1}). For typical non-ICL studies, this object detection scheme would be considered unusually aggressive because it finds sources whose S/N is as low as $\mytilde$1.7. A majority of the sources near this detection limit are spurious. However, this allows us to greatly reduce the impact of extremely faint sources that otherwise would masquerade as ICL. In addition, this low threshold allows us to obtain extended masking regions in the outskirts of objects. We let SExtractor convolve the WFC3 image with the FWHM$=$3 pixels Gaussian filter to optimize the detection of these faint features (Dalcanton et al. 1997). 

To determine the local sky background for object detection, we set {\tt BACK\_SIZE=128} in the SExtractor parameter; this size corresponds to 54 kpc at the cluster redshift, which is much greater than the average size of the galaxies in the field (see \textsection\ref{section_results}) and thus minimizes the impact of the unmasked outer regions of large objects in the local sky estimation. Increasing the block size beyond {\tt BACK\_SIZE=128} hampers us from correctly capturing the spatial variation of the sky. We use the resulting segmentation map obtained in this way as the masking image. If we had used the detection scheme {\tt DETECT\_MINAREA=5} and {\tt DETECT\_THRESH=1.5}, which is a more conventional setup corresponding to a minimum S/N of 3, we should have extended SExtractor's semi-major and -minor axes $\mytilde$5 times in order to match the masking region produced in the first setup.

\begin{figure}[!htb]
\centering
\includegraphics[width=8cm]{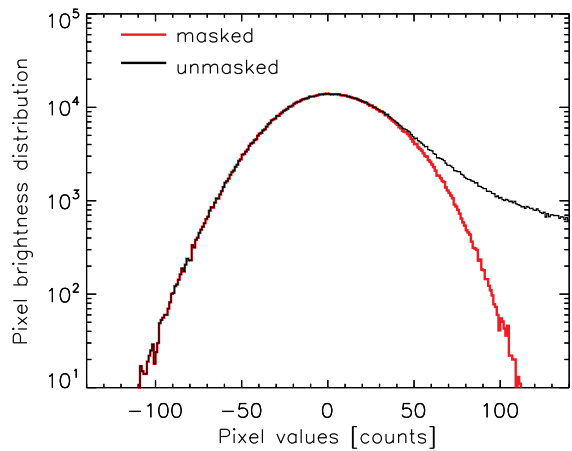}
\caption{Pixel intensity distributions of unmasked (black) and masked (red) images at $r= 36-48 \arcsec$ (300$-$400 kpc) in the F140W band. The skewness of the red curve is $-$0.001. We interpret this small skewness as indicating that our masking scheme sufficiently suppresses the contribution from extended diffuse wings of galaxy light profiles.}
\label{fig:px_dist}
\end{figure}

It is worth noting that we run SExtractor without deblending because incorrect deblending is common near very bright objects and this would prevent us from obtaining correct masking regions. Sometimes in the border region between objects, our object detection scheme makes isolated unmasked pixels surrounded by a number of masked pixels. In these cases, we manually fix the problem by masking out those isolated pixels. The importance of the proper masking scheme is illustrated in Figure~\ref{fig:px_dist}, where we display the histogram of pixel values in the F140W band at $300<\mbox{r}<400$ kpc (36--48 arcsec). This is the region where the diffuse light profile starts to flatten (i.e., we do not measure any excess diffuse light). If the applied masking area is not sufficiently large, the histogram becomes skewed to the right because of the diffuse light coming from the unmasked outer wings of objects. Using the following definition of skewness: 
\begin{equation}
Skewness = \frac{1}{N}\sum_{i=1}^{N}\left (\frac{x_{i}-\mu}{\sigma} \right )^3
\end{equation}
where $\mu$ is the mean and $\sigma$ is the standard deviation of the N pixel values, we find that the distribution after the correction has a skewness of $-$0.001, which is negligibly small.

\begin{figure*}[!htb]
\centering
\includegraphics[width=18cm]{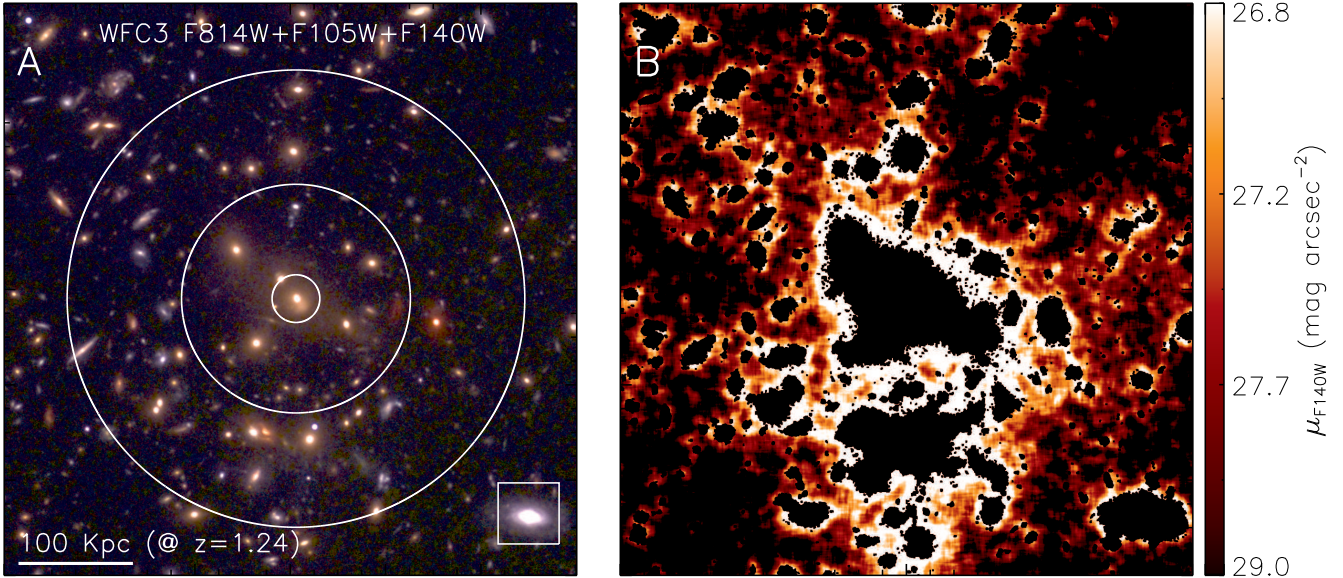}
\caption{Two-dimensional ICL detection. (A) HST WFC3 color composite image of MOO~J1014+0038. We use blue, green, and red to represent the intensities in the F814W, F105W, and F140W filters, respectively.
We show the central 1$\arcmin \times 1 \arcmin$ (500 kpc $\times$ 500 kpc) region. The three circles are centered on the BCG and have radii of 20, 100, and 200 kpc. The square in the lower right corner indicates the mesh size ($6.4\arcsec \times 6.4 \arcsec$) used for our local background estimation. (B) Diffuse light map in F140W. The map is created first by masking objects with a detection threshold of $\mytilde1\sigma$ above the local sky background and then by applying a $1\arcsec \times 1 \arcsec$ box median smoothing. The surface brightness threshold $\mu_{F140W}=29.0$ mag arcsec$^{-2}$ corresponds to $1\sigma$ sky above the background level. The diffuse light detected in the central region ($\mu_{F140W}=26.8$ mag arcsec$^{-2}$, $7\sigma$ sky) extends to $\mytilde 24\arcsec$ $(\mytilde$200 kpc) from the BCG. North is up and east is to the left.  }  
\label{fig:ICL_2D}
\end{figure*}

\section{Intracluster Light Measurements} \label{section_results}

\begin{figure*}[!htb]
\centering
\includegraphics[width=\textwidth]{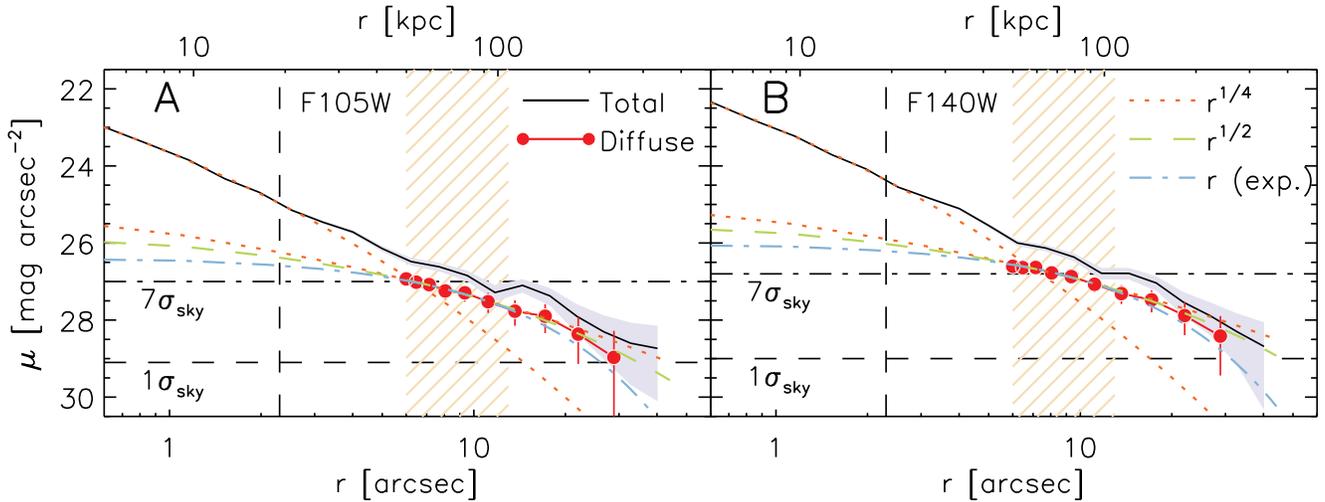}
\caption{Radial surface brightness (SB) profiles in F105W (A) and F140W (B). We plot azimuthally-averaged SB profiles of the total and diffuse (before and after masking out all objects) light in the cluster field as a function of radius. Black solid line and shaded region represent the SB of the total cluster light and the  errors, respectively. The dotted line overlapping with the solid line shows the best-fit de Vaucouleurs profile ($r^{1/4}$) for the inner regions from $0.9\arcsec$ to $2.3\arcsec$ (vertical dashed line) where the BCG is dominant in the total SB (see the innermost circle in Fig. 5A). Red filled circles illustrate the SB of the diffuse light. The error bars display the  uncertainties computed by combining the photon noise, residual flat-fielding systematics, and errors in background level determination. Orange dotted, green dashed, and blue dot-dashed lines represent the best-fit $r^{1/4}$, $r^{1/2}$, and exponential profiles for the $r=6\arcsec-13\arcsec$ region (orange hatched), respectively. The inner radius was determined by locating the radial bin where unmasked pixels start to dominate. The outer radius was chosen by determining the location where the signal-to-noise ratio decreases to 5. The horizontal dot-dashed and dashed lines indicate $7\sigma_{sky}$ and $1\sigma_{sky}$  levels above the background, respectively.}
\label{fig:ICL_profile}
\end{figure*}

\subsection{Feasibility of ICL Detection}

Typically, ICL in nearby clusters is studied below a surface brightness threshold of $\mytilde$25 (26) mag arcsec$^{-2}$ in the rest-frame optical B (R) band (Krick \& Bernstein 2007; Burke et al. 2015). To assess the feasibility of detecting ICL for \MOO~ with the current data, we performed the following feasibility test by estimating expected surface brightness threshold at $z=1.24$.

To start with, we converted the above B (R) band surface brightness thresholds into the equivalent values for F105W (F140W) at the cluster redshift first by adding cosmological surface brightness dimming ($1+z)^{4}$ and then by taking into account the stellar population evolution and the passband shift between the observed F105W (F140W) and rest-frame B (R) filters. We used the stellar population synthesis models of Bruzual \& Charlot (2003), assuming a simple stellar population with the formation redshift of $z_{f}=3$, the solar metallicity, and the Chabrier (2003) initial mass function. The corresponding surface brightness threshold turns out to be $\mytilde$27.3 (28.6) mag arcsec$^{-2}$ for the F105W (F140W) filter.

Thus, with our surface brightness limit ($\mytilde$29 mag arcsec$^{-2}$) derived from the background level uncertainty and $\mytilde$0.4 kpc spatial resolution, we are assured that the HST WFC3 images provide a clear two-dimensional ICL distribution extended to $\mytilde$200 kpc from the BCG as shown in Figure~\ref{fig:ICL_2D}. 

\subsection{Surface Brightness Measurement}
\label{section_total_diffuse}
We derived radial surface brightness [$\mu$ (azimuthally-averaged surface brightness)] profiles for the ICL as a function of radius from the center of the cluster (hereafter we adopt the position of the BCG as the center) (Figure~\ref{fig:ICL_profile}). Below, we refer to our surface photometry result as `total' (`diffuse') when obtained before (after) masking out all luminous objects. For the total light, the BCG is dominant at $r<2.3\arcsec$ ($\mytilde$20 kpc) and the radial surface brightness profile is well described by the de Vaucouleurs ($r^{1/4}$) profile. On the other hand, at $r>6\arcsec$ ($\mytilde$50 kpc) the diffuse light is better approximated by the exponential ($r$) profile, being less concentrated in the center. Our result suggests that the sum of the two components (the inner $r^{1/4}$  and outer $r^1$ profiles) might provide a reasonable fit to the combination of the BCG and ICL surface brightness profiles, which agrees with previous findings at $z<0.5$ (Gonzalez et al. 2005; Presotto et al. 2014), but is the first detection at $\mbox{z}>1$. This BCG+ICL decomposition of the total surface brightness profile suggests that the two components should be physically distinct and the outer light belongs to the intracluster stars rather than to the BCG (Cooper et al. 2015).

\begin{figure*}[!htb]
\centering
\includegraphics[width=\textwidth]{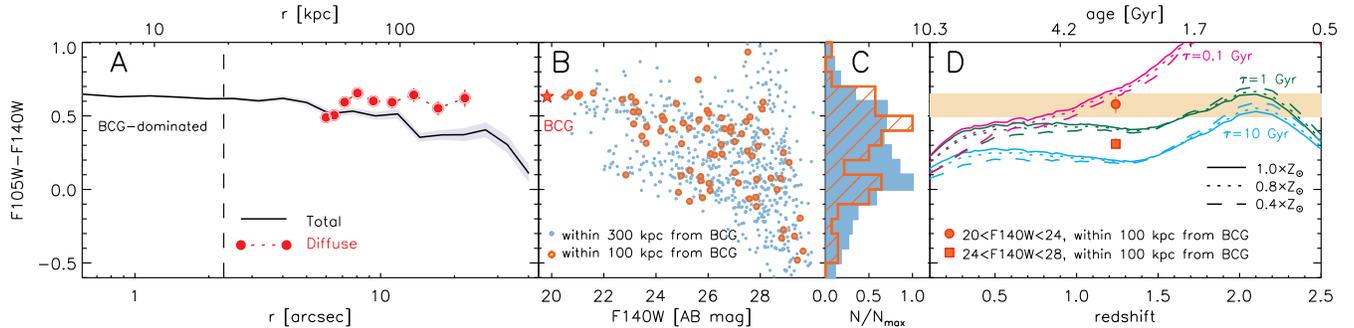}
\caption{Radial color profiles in the MOO J1014+0038 field. (A) F105W--F140W color profile of the total (black solid line) and diffuse (red dotted line) light. The shaded region and error bars include photon noise and residual flat-fielding errors. Although the total errors (dominated by our background level uncertainties) are much larger ($\mytilde0.3$ mag at $r=6-13\arcsec$), they do not affect our measurement of the relative color variation. (B) Color versus magnitude and (C) distribution of color for galaxies. Blue circles and histogram represent the measurement for galaxies within  300 kpc whereas orange circles and histogram for galaxies within  100 kpc. The star symbol indicates the observed color of the BCG. Note that the color of the diffuse light is consistent with those of the BCG and red bright galaxies near the BCG. (D) Evolution of F105W--F140W (apparent) colors of model galaxies at the formation redshift $z=3$. Magenta, green, and blue colors represent the exponentially decaying star formation model with $\tau=$ 0.1, 1, and 10 Gyr, respectively. Solid, dotted, and dashed lines show the solar, 80\% solar, and 40\% solar metallicities, respectively. Circle and square display the median F105W--F140W colors of bright ($20<\mbox{F140W}<24$) and faint ($24<\mbox{F140W}<28$) galaxies within  r=100 kpc from the BCG, respectively. The error bars show the 1$\sigma$ confidence interval estimated from bootstrapping. The shaded regions show the color distribution of the diffuse light out to $\mytilde$100 kpc from the BCG, where we detect the surface brightness at the 5$\sigma$ level in both filters.}
\end{figure*}
\label{fig:ICL_color_profile}

\subsection{Color Profiles}
It is well known that the optical color distribution of galaxies is bimodal with quiescent, bright galaxies populating a narrow red sequence and star-forming, faint galaxies forming a wide blue cloud in the color magnitude diagram (e.g., Strateva et al. 2001; Ko et al. 2013). Thus, the ICL color measurement allows us to constrain the progenitor galaxies and even the timescale of the ICL formation when a stellar population synthesis model is assumed. We created color profiles for the total and diffuse light by subtracting the surface brightness profiles at the same radial bin in two bands (Fig. 7A). We used a common mask for the F105W and F140W filters because any inconsistency in masking between different filters can cause bias in color profile measurement. This common mask was created by combining the F105W and F140W masks. However, even if we use the same mask, it is still possible that 
filter-dependent gradients of PSFs and/or galaxy profiles can cause
the amount of light spilled outside the masking area to be different. We believe that this is not a concern in our case 1) because the PSF profiles of F105W and F140W are similar in size and 2) because we took care so that our masking size is sufficiently large as explained in \textsection\ref{section_masking}.

The colors of the BCG and the bright galaxies located in the central region are consistent with that of the diffuse light out to $r\mytilde$200 kpc. This might suggest that the diffuse light originates from the BCG or/and bright red galaxies in the central region of the cluster. The total light gradually becomes bluer beyond $r\mytilde$50 kpc. This change is expected from the radial color variation of galaxies (Fig. 7B). On the other hand, this trend is not observed for the diffuse light.

Our finding is consistent with the pattern observed in previous studies of clusters at $\mbox{z}=0.2-0.4$ (Zibetti et al. 2005; Jee 2010) that the average color of the ICL is similar to that of the bright, red galaxies (including BCGs). Furthermore, the relatively flat radial color profile may indicate that the ICL is of a simple stellar population with a similar metallicity (e.g., Franx et al. 1990) or is a consequence of mixed stellar populations, such as flatten color gradients of bright cluster galaxies by major merging events (e.g., Ko \& Im 2005).

To constrain the formation epoch of the ICL, we consider the evolution of the intrinsic galaxy color by applying a stellar population synthesis model (Fig. 7D). We adopt an exponentially declining star formation history characterized by the e-folding time $\tau$. Figure 7D shows the F105W$-$F140W colors of the central bright and faint galaxies compared with the predicted colors of model galaxies formed at $z_{f}=3$ with $\tau=$ 0.1, 1, and 10 Gyrs. For the BCG and bright galaxies within 100 kpc, the comparison between observation and model indicates that they formed during a short period and then passively evolved to the cluster redshift. Furthermore, the models with $\tau=0.1-1.0$ predict that the stellar color is 0.5$-$0.7 at $z_{f}=2-6$, corresponding to the epochs 2$-$4 Gyrs earlier than the cluster redshift. Therefore, one possible explanation is that those stars in the ICL could have formed at $z\mytilde$2 or earlier and might have migrated to the ICL during the period from their formation to the cluster redshift ($z=1.24$), mainly through the interaction between the BCG and bright central galaxies. Our finding thus supports the hypothesis that the ICL is already in place long before $\mbox{z}=1$. Note that this early ICL formation is not favored in previous theoretical (Rudick et al. 2011; Contini et al. 2014; Cooper et al. 2015) and observational (Zhang et al. 2016; Montes \& Trujillo 2018) studies, which support a late ($\mbox{z}<1$) ICL formation scenario.

\subsection{ICL Contribution to Cluster Light}

\begin{figure}
\centering
\includegraphics[width=8cm]{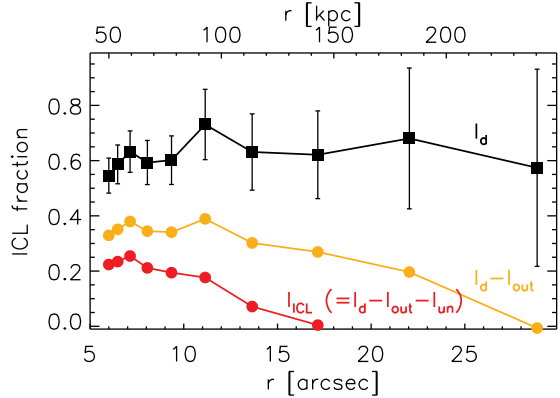}
\caption{ICL fraction. Black line displays the ratio of diffuse light to total cluster light in F140W. Error bars represent the 1$\sigma$ uncertainties including both statistical and systematic errors such as flat-fielding errors and background level uncertainties. Orange line shows the result when masking incompleteness is corrected for. Red line includes the corrections from both this masking incompleteness and the contamination from undetected diffuse, faint galaxies.}
\label{fig:ICL_fraction}
\end{figure}

A further important constraint is possible when one investigates the ICL fraction as a function of the cluster-centric radius. Traditionally, the ICL fraction is defined as the ratio of the diffuse light to the total cluster light at a certain reference radius. However, several ambiguities are present. First, it is difficult to accurately measure the total cluster light unless one has access to a highly complete spectroscopic catalog of the cluster field. Because our study is not immune to this issue either, our total luminosity measurement is subject to overestimation. 
Thus, our ICL fraction  presented here should correspond to a lower limit; some previous studies suggest that contamination from non-cluster galaxies is not a serious problem in measuring ICL fractions (e.g., Zibetti et al. 2005; Burke et al. 2012).
Second, the diffuse light measurement is contaminated by contributions from unmasked galaxy/PSF profiles and undetected low surface brightness objects. Although many studies in the past often ignored this contamination, a growing number of recent papers show that the contamination is not negligible and must be considered as a critical limiting factor. To address the issue, we decompose the diffuse light ($I_{d}$) into three main components: $I_{d}=I_{ICL}+I_{out}+I_{un}$, where $I_{out}$ and $I_{un}$ are the light components from the pixels outside the masking area (but still belonging to objects) and undetected faint, diffuse galaxies, respectively; we
present descriptions on how we measure these quantities in \textsection\ref{section_mask_incompleteness}~and~\ref{section_diffuse_galaxy}.

By quantifying the latter two components, we estimate the ICL contribution to the total cluster light. In Figure~\ref{fig:ICL_fraction}, we show the diffuse ($I_{d}$), $I_{out}$-subtracted ($I_{d}-I_{out}$), and $I_{out}$- and $I_{un}$-subtracted ($I_{d}-I_{out}-I_{un}=I_{ICL}$) fractions. According to our conservative estimation, the $I_{out}$ component is found to be as high as 40\%-80\% of the diffuse light ($I_{d}$) within the projected distance of $r\mytilde$200 kpc. Beyond $\mytilde$150 kpc the ICL fraction converges to zero, which implies that the diffuse light ($I_{d}$) there is dominated by $I_{out}+I_{un}$. Despite this conservative approach, it is remarkable that still more than 10\% of the total light is contributed by the ICL at $\mbox{r}\leq$110 kpc. This ICL fraction that we determine from MOO~J1014+0038 at $\mbox{z}=1.24$ is comparable to those of Virgo (Mihos et al. 2017) and other nearby clusters (Feldmeier et al. 2004).

One caveat is that observational methods of measuring the total cluster light and ICL are different between Virgo (and other nearby clusters) and our cluster. Hence, a comparison of their ICL fractions on an equal footing is  difficult. Nevertheless, it is worthy to note that Mihos et al. (2017) presented a rough estimate of the total ICL fraction of Virgo to be 7\%$-$15\% (by tracing the luminosity of the observed tidal features), based on the simulations of Rudick et al. (2011), where the cluster luminosity below the threshold $\mytilde$26.5 mag arcsec$^{-2}$ in $V$ band is mainly contributed by stripped (unbound) stars.

Our result supports the significant presence of the ICL long before $z\sim1$, when most of the present-day BCG mass was assembled (e.g., Lidman et al. 2013). 
Burke et al. (2012) found that the ICL constitutes 1\%$-$4\% of the total cluster light within a radius $R_{500}$ at the surface brightness limit of $\mytilde$23 mag arcsec$^{-2}$ in J band, analyzing six massive galaxy clusters at 0.8$\leq\mbox{z}\leq$1.2. From their comparison with low-redshift ICL studies using a similar surface brightness threshold, they suggested that the fraction of the ICL has increased by a factor of 2$-$4 since $z\sim1$. However, if \MOO~ is representative of typical clusters at $z\sim1$, our result does not support such a large growth in ICL fraction.

\subsubsection{Mask incompleteness} \label{section_mask_incompleteness}
Although we take care in determine masking sizes conservatively (\textsection\ref{section_masking}), inevitably a substantial fraction of our surface brightness profile presented in \textsection\ref{section_total_diffuse} is contributed by unmasked object profiles. In principle, one can address the issue by modeling and subtracting object profiles individually. However, the exact SB profiles of individual galaxies vary quite significantly and thus are hard to model (especially for those in the central region where ICL is dense). In this paper, instead of modeling and subtracting galaxy profiles object by object, we used a statistical approach to estimate the contribution as below.

Assuming that we know the average surface brightness profile of galaxies $I(r)$, we can model the amount of the average surface brightness $I_{out}$ outside masks for a given total diffuse surface brightness level $I_d$ as
\begin{equation}
I_{out} = \frac{\int_{r_{d}}^{r_{0}} I(r)2\pi r dr}{\int_{r_{d}}^{r_{0}} 2\pi r dr}, \label{eqn_I_out}
\end{equation}
\noindent
where $r_{0}$ and $r_{d}$ are the radii at which the surface brightness profiles $I(r)$ become our detection threshold (29.0 mag arcsec$^{-2}$) and the total diffuse light brightness ($I_d$), respectively.  Because we let the upper limit of the surface brightness be equal to the total surface brightness level $I_d$ in Equation~\ref{eqn_I_out}, this estimation is conservative and may lead to overestimation (the actual maximum surface brightness level of unmasked object profiles contributing to $I_d$ may be lower).

In the above we make two assumptions. The first one is that the distribution of galaxy light is roughly independent of radial bins at $r>50$ kpc from the BCG. The second assumption is that the average SB profile of galaxies at each radial bin is represented by $I(r)$, which is constructed from the bright (20$-$22 mag in F140W) and isolated (no sources brighter than 22 mag in F140W within a radius of 4 arcsec) galaxies in the $42\arcsec-60\arcsec$ annulus from the BCG, where the ICL contribution to the surface brightness of the galaxy wings is negligible.

The surface brightness measurement of the galaxy wings obtained in this way already include the contribution from the point spread function (PSF) wings. Also, no stars with significant brightness are present within 200 kpc from the BCG, where we measure the ICL properties. Therefore, it is not necessary to separately take into account the PSF wing contribution to the ICL measurement in our analysis. 

\subsubsection{Undetected diffuse, faint galaxies} \label{section_diffuse_galaxy}
The limiting magnitude of the F140W image ($\mytilde$28.5 mag at the 5$\sigma$ limit) approximately corresponds to the R-band absolute magnitude $M_{R}\mytilde-$14.1 mag; we adopt a distance modulus of 44.73 and correct for the passband shift between the observed F140W and rest-frame R filters while considering stellar population evolutions (a simple stellar population model formed at $z=3$). This limiting magnitude is comparable to the magnitude of the faintest normal galaxies in the Coma cluster (Mobasher et al. 2001), but not sufficiently low to enable detection of ultra-diffuse galaxies (UDGs) (van Dokkum et al. 2015; Koda et al. 2015; Yagi et al. 2016). We regard the UDG population as one of the diffuse light components in the current study because several lines of evidence indicate that a great number of UDGs might also present in the cluster at $\mbox{z}>1$ (e.g., van der Burg et al. 2017).

To make a quantitative estimation, we adopt the result (Koda et al. 2015) of the Coma cluster, where they determine the mean central surface brightness to be $\mu_{0}=24.6$ mag arcsec$^{-2}$ from their 854 UDGs in the magnitude range $-16<M_{R}<-12$. This central surface brightness corresponds to the observed surface brightness $\mu_{0}\sim$27.2 mag arcsec$^{-2}$ in F140W. This central surface brightness can be converted to the mean surface brightness within the effective radius as $<\mu(r_{e})> \sim \mu_{0}+1.12$  by assuming an exponential profile because the average S{\'e}rsic index is 0.9 for the Coma UDGs (Koda et al. 2015). The resulting $<\mu(r_{e})>\sim$28.3 mag arcsec$^{-2}$ in F140W is adopted as undetected diffuse light component $I_{un}$. 

It is possible that we slightly overestimate $I_{un}$ because of the following two reasons. The first one is that the central surface brightness $\mu_{0}\sim27.2$ is higher than our object detection threshold and the cores of the UDGs may have already been removed in our measurement of the diffuse light. The second one is that in our analysis we consider $I_{un}$ to be constant as if the UDGs are uniformly distributed throughout the cluster even near the cluster center (Mihos et al. 2015; Mu{\~n}oz et al. 2015; Yagi et al. 2016). However, it is worth noting that theoretical studies expect that UDGs having much younger ages fall into the clusters later than typical cluster dwarf galaxies (Rong et al. 2017). This implies that the UDGs might be less populated in the core regions of the current cluster at $\mbox{z=1.24}$ than nearby clusters. Thus, our subtraction of the UDG contribution from the diffuse light should lead to conservative values for the ICL fraction.

\section{Conclusions} \label{section_conclusions}
We have presented the ICL study of MOO~J1014+0038 at $z=1.24$. This is the highest-redshift cluster to date, for which we measure the two-dimensional distribution, as well as the radial color profile. The high-quality HST/WFC3 near-IR imaging data enables us to reach a very low surface brightness threshold ($\sim$29 mag arcsec$^{-2}$) and obtain a clear two-dimensional ICL map out to $\mytilde200$ kpc from the cluster BCG.

We find that the ICL color is consistent with that of the bright, red cluster galaxies. However, unlike the radial color variation of galaxies, we do not detect any significant radial dependence of the ICL color. Using simple stellar population synthesis with an exponentially decaying star formation model, we estimate that the ICL stars had formed at $z\sim2$ or earlier. 

When estimating the ICL fraction, we take into account the contributions from the pixels outside our masking regions and from undetected faint, diffuse galaxies. In our most conservative case, the unmasked pixels contribute as much as $\mytilde80$\% of the total diffuse light within $r=200$ kpc. It is remarkable that despite this conservative analysis, the integrated ICL fraction still exceeds $\mytilde10$\% of the total cluster light at $r<200$ kpc, comparable to measurements in low-redshift clusters.

Currently, two dominant physical mechanisms have been proposed to explain the formation of ICL: tidal stripping of the outskirts of infalling/satellite galaxies (e.g., Contini et al. 2014; Cooper et al. 2015) and violent mergers of cluster members during the formation of the BCG (e.g., Murante et al. 2007; Conroy et al. 2007). Both mechanisms may be at work. However, the dominance may be a function of time during the hierarchical growth of the cluster.

The time difference between the cluster redshift ($z=1.24$) and the formation epoch $z=2$ (3) is $\mytilde$1.7 (2.8) Gyrs, during which the cluster galaxies can traverse the cluster only once or twice (assuming a free fall time for a massive cluster with a radius of $\mytilde$1 Mpc and a velocity dispersion of $\mytilde1000~\mbox{km}~\mbox{s}^{-1}$). Thus, if the ICL formation is an ongoing process and predominated by the stripping of the outskirts of infalling/satellite galaxies, we should be able to observe the evolution of ICL fraction between $z=0$ and 1. However, the presence of the significant ICL fraction at $z=1.24$ strongly supports the paradigm that the dominant process for the ICL production is linked to the BCG formation, although we need to perform further analysis on more galaxy clusters at $z>1$ to confirm that the cluster sample studied here is not exceptional.

\acknowledgments
{We thank the anonymous referee for useful comments that greatly improved this paper. M.J.J. acknowledges support for the current research from the National Research Foundation of Korea under the program 2017R1A2B2004644 and 2017R1A4A1015178.}

\end{document}